\documentclass[preprint,12pt]{elsarticle}
\usepackage{amsmath}
\usepackage{amsfonts}
\usepackage{amssymb}
\usepackage{amsthm}
\usepackage{graphicx}
\usepackage{color}

\usepackage[normalem]{ulem}

\def\x{{\mathbf x}}
\def\y{{\mathbf y}}
\def\a{{\mathbf a}}
\def\w{{\mathbf w}}
\def\b{{\mathbf b}}

\def\A{{\mathbf A}}
\def\I{{\mathbf I}}
\def\P{{\mathbf P}}
\def\R{{\mathbf R}}
\def\B{{\mathbf B}}
\def\E{{\mathbf E}}
\def\W{{\mathbf W}}

\newcommand{\bfPi}{{\mbox{\boldmath $\Pi$}}}

\newif\ifpfig
\pfigtrue

\begin{document}
\begin{frontmatter}



\title{A Search-free DOA Estimation Algorithm for Coprime Arrays}


\author{Zhiyuan Weng and Petar M. Djuri\'{c}}

\address{Stony Brook University, Stony Brook, New York 11794}

\begin{abstract}
Recently, coprime arrays have been in the focus of research because of their potential in exploiting redundancy in spanning large apertures with fewer elements than suggested by theory.
A coprime array consists of two uniform linear subarrays with inter-element spacings $M\lambda/2$ and $N\lambda/2$, where $M$ and $N$ are coprime integers and $\lambda$ is the wavelength of the signal.
In this paper, we propose a fast search-free method for direction-of-arrival (DOA) estimation with coprime arrays. It is based on the use of methods that operate on the uniform linear subarrays of the coprime array and that enjoy many processing advantages. We first estimate the DOAs for each uniform linear subarray separately and then combine the estimates from the subarrays. For combining the estimates, we propose a method that projects the estimated point in the two-dimensional plane onto one-dimensional line segments that correspond to the entire angular domain. By doing so, we avoid the search step and consequently, we greatly reduce the computational complexity of the method. We demonstrate the performance of the method with computer simulations and compare it with that of the FD-root MUSIC method.
\end{abstract}


\end{frontmatter}

\section{Introduction}
Direction-of-arrival (DOA) estimation using sensor arrays is a problem that is frequently encountered in many engineering areas including radar, sonar, and wireless communication, and it has been studied for several decades. Recently, the notion of coprime array signal processing has emerged as an area of interest where, as the name suggests, the processing is applied to signals acquired by coprime arrays \cite{pal2011coprime,vaidyanathan2011sparse,vaidyanathan2011theory,Negusse20122254}. In general, it has been shown that coprime sampling allows us to sample a signal sparsely and estimate parameters of signals at higher resolutions \cite{vaidyanathan2011sparse}. The parameters can be the frequencies of temporal signals or the DOAs of spatial signals. 
By sparse sampling, one can reduce the rate of analog-to-digital converter for sampling in the temporal domain \cite{Negusse20122254} or reduce the number of array sensor elements for sampling in the spatial domain \cite{pal2011coprime}. This entails that systems where coprime sampling is applied can have lower cost than systems with Nyquist sampling, yet without degrading the performance of the latter. For example, the cost of ADC usually grows exponentially with the bandwidth or the sampling rate. By using coprime sampling, one can reduce the requirements on the analog front-end by significantly lowering the sampling rate \cite{Negusse20122254}.

In  general, the idea behind coprime arrays is to extend the concept of minimally redundant sensor arrays and span large apertures by using far fewer elements than prescribed by textbook antenna theory. A coprime array can be constructed with two uniform linear arrays with $M$ and $N$ elements with inter-element spacings $M\lambda/2$ and $N\lambda/2$, respectively, and where $M$ and $N$ are co-prime integers, i.e., integers whose greatest common divisor is one. It has been shown that in the presence of spatially stationary fields, one can match the information extracted with such arrays to that of fully populated arrays with $MN$ elements \cite{vaidyanathan2011sparse}. As already pointed out, a very important payoff for the use of coprime arrays is simplified and reduced cost array designs. Instead of having an array with $MN$ elements, one can use an array with $M+N-1$ elements (where one element is shared).

The reduced cost of coprime arrays represents a strong motivation for studying DOA methods that can process signals for such arrays. An obvious approach is to consider methods that are developed to operate on arbitrary geometry arrays. In this class there are several popular DOA estimation methods including beamspan \cite{mayhan1987spatial}, Capon \cite{capon1969high}, MUSIC \cite{schmidt1986multiple} and several versions of maximum likelihood estimators \cite{stoica1990performance}. For finding the optimal estimate, however, they all require a computationally expensive search step in a nonconvex space. We note that for uniform linear arrays (ULAs), the array response vector has a Vandermonde form, which allows the search step to be replaced by polynomial rooting. These so called \emph{search-free} algorithms include IQML \cite{bresler1986exact}, MODE (Method of Direction Estimation) \cite{stoica1990maximum}, root-MUSIC \cite{rao1989performance} and ESPRIT \cite{roy1989esprit}. An interesting alternative to the methods for arbitrary arrays are the approaches that transform the arbitrary arrays to equivalent ULAs so that one can take the advantage of efficient algorithms for ULAs, e.g., array interpolation  \cite{friedlander1992direction}, manifold separation \cite{belloni2007doa} and Fourier-Domain (FD) root-MUSIC \cite{rubsamen2009direction}. These approaches basically try to approximate the steering vectors by virtual ULAs. Their drawback is that for achieving satisfactory performance, they require a large number of virtual arrays, which entails increased complexity. Furthermore, they suffer fixed error at high signal-to-noise ratios (SNRs) and thus, do not converge to the Cram\'{e}r-Rao bound as the SNR increases.


The DOA estimation with ULAs has been thoroughly studied, but there has not been much work for designing special algorithms that take advantage of the \emph{partial} uniform linear structure of coprime arrays. In \cite{pal2011coprime}, the authors apply the MUSIC algorithm to coprime processing, where they mainly consider the problem from the perspective of degree of freedom, i.e., how many signals the coprime array can identify. To achieve additional degree of freedom with coprime arrays, that is, to enhance the rank of the covariance matrix of the received signal, they propose the use of spatial smoothing. Nevertheless, the proposed method is not search-free, and its accuracy and computational complexity have not been investigated in detail. In \cite{Negusse20122254}, where processing of time series is addressed, the authors use an iterative method to search for the optimal frequency. Although it is search-free, it does not use the coprime property and it cannot handle the case of multiple signals.

In this paper, we address the problem of DOA estimation with coprime arrays with the emphasis on reduced computational complexity while preserving estimation accuracy. We propose a search-free method that exploits the uniform linear structure of the subarrays of the coprime arrays. Considering that the DOA estimation using ULAs is fast and accurate, we first estimate the DOA for each subarray separately. Since the inter-element distance for each subarray is larger than half-wavelength of the signal, ambiguity is present and as a result, the DOAs cannot be uniquely determined. To resolve the ambiguity, we use a projection-like method that combines the estimates from different subarrays, where the correctness of the estimate is guaranteed as a consequence of the coprimeness.

The paper is organized as follows. In Section 2, we introduce the model and briefly review the DOA estimation with ULAs. The proposed algorithm is discussed in detail in Section 3 and the numerical experiments are presented in Section 4. Section 5 provides final remarks.\footnote{An earlier version of this paper can be found at http://arxiv.org/abs/1301.4155.}

We use the following notation: $(\cdot)^{\top}$ and $(\cdot)^{{\rm H}}$ denote the transpose and the conjugate transpose, respectively; $\mathbb{C}$ refers to the complex space; $\delta_{k,l}$ signifies the Kronecker delta; $\mathbb{E}(\cdot)$ stands for the expectation operator; and $\textrm{tr}\{\bf A\}$ is the trace of the matrix ${\bf A}$.

\section{DOA estimation for ULA}
The problem of finding the DOAs of $D$ narrowband plane waves impinging on a ULA of $L$ sensors can be modeled as follows \cite{krim1996two}:
\begin{align}
\y(k)=\A\x(k)+\w(k),~~~k=1,\cdots,K,
\end{align}
where $\y(k)\in\mathbb{C}^{L\times 1}$ is the signal received by the sensors at the $k$th time slot;
$K$ is the number of snapshots;
$\x(k)\in\mathbb{C}^{D\times 1}$ contains the complex envelopes of the emitter signals;
$\w(k)$ is a noise process;
and
\begin{align}
\A=[\a_1,\cdots,\a_D]\in\mathbb{C}^{L\times D}
\end{align}
with
\begin{align}\label{steeringvector}
\a_d=[1, \exp(jp_1\psi_d), \cdots, \exp(jp_{L-1}\psi_d)]^{\top} \in\mathbb{C}^{L\times 1}\nonumber\\
~~~~~~~~~~~~~~~~~~~~~~~~~~~d=1,\cdots,D;
\end{align} $p_l\lambda/2$ is the distance from the $l$th sensor to the reference point; $\psi_d$ is the DOA of our interest. The number of sources, $D$, is assumed to be known. Thus, $\Psi=[\psi_1,\cdots,\psi_D]$ is the vector of DOAs we wish to estimate. We assume the signal $\x(k)$ and the noise $\w(k)$ are independent zero-mean complex Gaussian random processes with the following moments:
\begin{align}
\mathbb{E}[\x(k)\x^H(l)]&=\P\delta_{k,l},~~~&\mathbb{E}[\x(k)\x^T(l)]&=\boldsymbol{0}, k\neq l\\
\mathbb{E}[\w(k)\w^H(l)]&=\sigma^2\I\delta_{k,l},~~~&\mathbb{E}[\w(k)\w^T(l)]&=\boldsymbol{0}, k\neq l
\end{align}
with $\I$ being the identity matrix, $\P$, the signal covariance matrix, and $\sigma^2$, the noise power.
The covariance matrix of the received signal $\y(k)$ can be written as
\begin{align}
\R=\A\P\A^{\rm H}+\sigma^2\I.
\end{align}
We denote by $\hat{\R}$  the sample covariance matrix
\begin{align}\label{covR}
\hat{\R}=\dfrac{1}{K}\sum_{k=0}^{K-1}\y(k)\y^{\rm H}(k).
\end{align}
The eigendecomposition of $\hat{\R}$ in \eqref{covR} can be written as
\begin{align}
\hat{\R}=\hat{\E}_s\hat{\boldsymbol{\Lambda}}_s\hat{\E}_s^{\rm H}+\hat{\E}_n\hat{\boldsymbol{\Lambda}}_n\hat{\E}_n^{\rm H},
\end{align}
where
$\hat{\boldsymbol{\Lambda}}_s\in\mathbb{C}^{D\times D}$ and $\hat{\boldsymbol{\Lambda}}_n\in\mathbb{C}^{(L-D)\times (L-D)}$ are diagonal matrices that contain the eigenvalues of the signal and the noise subspaces, respectively, whereas $\hat{\E}_s\in\mathbb{C}^{L\times D}$ and $\hat{\E}_n\in\mathbb{C}^{L\times (L-D)}$ are composed of the eigenvectors of the signal and the noise subspaces, respectively.

For ULAs, the steering vector becomes
\begin{align}
\a_i(\psi_i) = [1,\exp(j\psi_i),\cdots,\exp(j(L-1)\psi_i)]^{\top}.
\end{align}
There exists a large number of fast algorithms for estimating DOAs with  ULAs, among which MODE is the leading candidate \cite{li1998comparative,van2002optimum}. Since we later use MODE as part of our algorithm, we review it here briefly. The MODE algorithm was originally proposed in \cite{stoica1990maximum}. It estimates the DOAs via the minimization of the following function:
\begin{align}\label{minobj}
\b=\underset{\b\in\mathbb{C}^{D\times 1}}{\arg\min}~\textrm{tr}\{\bfPi_{\B}\hat{\E}_s\W\hat{\E}^{\rm H}_s\}
\end{align}
where
\begin{align}
\mathbf{b}&=[b_0,b_1,\cdots,b_D]^\top,\\
\bfPi_{\B}&=\B({\B}^{\rm H}\B)^{-1}{\B}^{\rm H},\\
\W&=(\hat{\boldsymbol{\Lambda}}_s-\hat{\sigma}^2\I)^2\hat{\boldsymbol{\Lambda}}_s^{-1},\\
\hat{\sigma}^2&=\dfrac{1}{L-D}\textrm{tr}\{\hat{\boldsymbol{\Lambda}}_N\},
\end{align}
and $\B$ is a $L\times (L-D)$ Toeplitz matrix defined by
\begin{align}
\B^{\rm H} = \begin{bmatrix}
b_D & \cdots & b_0 & 0 & \cdots & 0\\
0 & b_D & \cdots & b_0 & \cdots & 0\\
\vdots & \ddots & \ddots & & \ddots & \vdots\\
0 & \cdots & 0 & b_D & \cdots & b_0
\end{bmatrix}.
\end{align}
Let $z=\exp(j\psi_i)$, and denote by $b(z)$ the complex-valued polynomial
\begin{align}
b(z)&=b_0z^D+b_1z^{D-1}+\cdots+b_D\\
&=b_0\prod_{i=1}^D(z-z_i).\label{poly}
\end{align}
The angles of the roots of the polynomial are the estimates of the DOAs. It was shown in \cite{stoica1990maximum} that the MODE algorithm is an asymptotically efficient estimator of the DOAs. {As per \cite{van2002optimum,li1998comparative}, convergence of MODE is guaranteed.} To minimize \eqref{minobj}, both iterative and non-iterative steps have been proposed. See \cite{li1998comparative, van2002optimum} for detail.

\section{DOA estimation for coprime arrays}\label{section3}
\ifpfig
\begin{figure}[h]
\centering
\includegraphics[scale=1.5, trim=240 615 450 90]{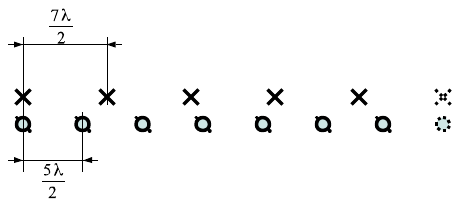}\\
\caption{a coprime array with two ULAs, where $M=7$ and $N=5$.}\label{fig1}
\end{figure}
\fi
We consider a coprime array with two uniform linear subarrays. We assume that the two subarrays have inter-element spacing $M\lambda/2$ and $N\lambda/2$, respectively, with $M$ and $N$ being coprime. Fig. \ref{fig1} shows the case for $N=7$ and $M=5$. Because the two subarrays share the first sensor, the total number of sensors $L$ is equal to $M+N-1$. The corresponding steering vector in \eqref{steeringvector} becomes
\begin{align}
\mathbf{a}_d=\left[1,e^{j\psi_dM},e^{j\psi_d2M},\cdots,e^{j\psi_d(N-1)M}\right.\nonumber\\
\left.e^{j\psi_dN},e^{j\psi_d2N},\cdots,e^{j\psi_d(M-1)N}\right]^{\top}.
\end{align}
For such arrays, the efficient methods for ULAs are not directly applicable. We can of course use the algorithms for arbitrary arrays, but they are slow and the uniform linear structure of the subarrays is wasted. Our objective is to develop an algorithm that is applicable to coprime arrays while at the same time enjoys the efficiency brought by the uniform linear structure of the subarrays. The proposed approach is composed of two steps. We first estimate the DOAs from each subarray separately, which is similar to the approach in \cite{Negusse20122254}. We then combine the two estimates in an innovative way.
\ifpfig
\begin{figure}[h]
\centering
\includegraphics[scale=1, trim=250 625 250 30]{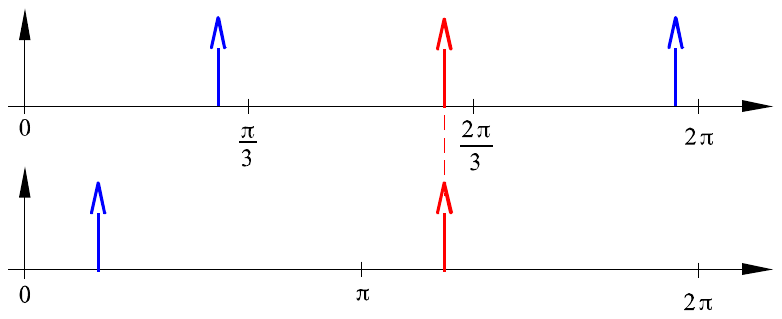}\\
\caption{Ambiguity on the angular domain, case 1.}\label{fig2}
\end{figure}
\fi
\ifpfig
\begin{figure}[h]
\centering
\includegraphics[scale=1, trim=250 625 250 40]{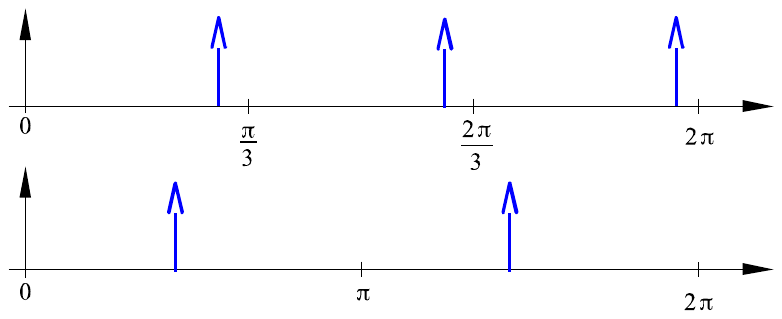}\\
\caption{Ambiguity on the angular domain, case 2.}\label{fig3}
\end{figure}
\fi
\subsection{DOA estimation using a single subarray}
For a subarray, the inter-element spacing is $N\lambda/2$ with $N>1$. This has the same effect as we undersample a signal by a factor of $N$ in the temporal domain. According to the basic sampling theorem, we will have ambiguities, or aliasing, in the angular domain. Specifically, if we use the same algorithm as described above, the polynomial in \eqref{poly} becomes
\begin{align}\label{poly2}
b'(z)=b'_0\prod_{i=1}^D(z^N-z'_i).
\end{align}
This polynomial has $N\times D$ roots instead of $D$. If $z_i$ is a root of \eqref{poly}, it is still a root of \eqref{poly2}. The problem is that aliasing occurs with the period of $2\pi/N$; $z_i\exp(jn\frac{2\pi}{N})$ for $n=1,\cdots,N-1$ are also the roots of \eqref{poly2}, as shown in Fig. \ref{fig2}. Therefore, we have no way of distinguishing the $z_i$ we want from the remaining $N-1$ roots.

Note that in the first step, it does not matter which algorithm we use. In the experiment, we selected MODE, which was shown to have the best performance for ULAs \cite{van2002optimum}.

\subsection{Combination of the estimates from subarrays}
The present ambiguities in the estimates from the two subarrays notwithstanding, we show that we are able to disambiguate them. For example, in the case of one source and no noise, in Fig. \ref{fig2}, we can see that the estimates from the two subarrays coincide at the red arrows. Therefore, we can be very confident that the position of the red arrow is the DOA. In the presence of noise, however, coincidences as the one in Fig. \ref{fig3} are unlikely, and consequently, it becomes difficult to tell the exact DOA. Furthermore, in the case with multiple sources, obtaining the DOAs in this way becomes impossible.

We propose a projection method that can uniquely determine the exact DOA. Recall that the aliasing period in the angular domain for the subarray is $2\pi/N$. Thus, we only need to take the values between $[-\pi,-\pi+2\pi/N]$. The ranges of the outputs of the two subarrays define a rectangle in the two-dimensional plane as shown in Fig. \ref{fig4}, where $N=2$ and $M=3$. It is not difficult to see that the 45 degree oblique line segments colored in blue correspond to the entire angular domain. The Chinese remainder theorem guarantees that the map is one-to-one and onto as a result of the coprimeness.  This is the key point of our algorithm. In Fig. \ref{fig4}, the entire angular domain consist of four line segments; L1 corresponds to $[-\pi, -\pi/3]$; L2, $[-\pi/3,0]$; L3, $[0, \pi/3]$; and L4, $[\pi/3, \pi]$. For general $M$ and $N$, there will be $M+N-1$ oblique line segments. Suppose the outputs of the first and the second subarray are $\psi^{(1)}$ and $\psi^{(2)}$, respectively. The two estimates specify a point in the plane. We project the point onto the oblique line segments that corresponds to the entire angular domain. Specifically, we seek the point on the oblique line segments that is nearest to that point specified by the two estimates. This ensures that the combination is optimal and the result is valid. For point \textbf{B} in Fig. \ref{fig4}, we can simply draw a line that is vertical to the oblique line and measure the distance from \textbf{B} to the intersections. For point \textbf{A}, however, one intersection falls outside the rectangle area and should be wrapped around. As a result, the intersection is actually on L3 in the figure. After the nearest point on the line segments is found, it is easy to calculate the corresponding DOA by solving a set of modular linear equations. As an alternative to solving the equations, one may have a precalculated lookup table with solutions of the modular equations.

The main difference between the proposed method and the one in \cite{Negusse20122254} is that here we take advantage of the coprimeness and use a projection process in the search for optimal DOAs, while the method in \cite{Negusse20122254} iteratively tries different combinations of the estimates and find the best one among them. Although solving the modular equations in our method requires an iterative process, it can be avoided by using a lookup table.

\subsection{Combination of the estimates in the case of multiple sources}
In the case of multiple sources, the method gets more complicated. Suppose that the outputs of the first subarray are $\psi_1^{(1)}$ and $\psi_2^{(1)}$ and the outputs of the second subarray, $\psi_1^{(2)}$ and $\psi_2^{(2)}$. We have no way of knowing which value corresponds to the first and which to the second target. Therefore, we have to check all the four points. In Fig. \ref{fig4}, $\psi_1^{(1)}$, $\psi_2^{(1)}$, $\psi_1^{(2)}$ and $\psi_2^{(2)}$ define four points on the two-dimensional plane, which are marked \textbf{A}, \textbf{B}, \textbf{C}, and \textbf{D}.
Since candidate points are obtained, we can readily evaluate the likelihood probability for these points and take the best $D$ ones.
\ifpfig
\begin{figure}[h]
\centering
\includegraphics[scale=1, trim=250 500 250 55]{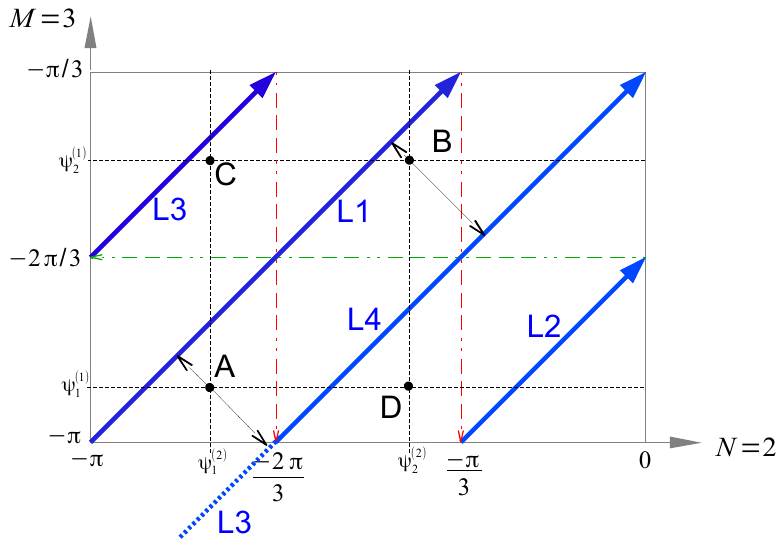}\\ 
\caption{Projecting the point specified by the two estimates onto the entire angular domain.}\label{fig4}
\end{figure}
\fi

In short, the proposed algorithm can be summarized as follows:
\begin{enumerate}
  \item Estimate the $D$ angles for each subarray.
  \item Plot the corresponding $D^2$ points on the 2-dimensional plane.
  \item Project these points to the nearest diagonal line segments.
  \item Map the projected $D^2$ points back onto the entire angular domain.
  \item Evaluate the likelihood probabilities at the $D^2$ points.
  \item Select $D$ points with the largest likelihood probabilities.
\end{enumerate}

\subsection{Example}
In this subsection, we provide a simple example to better illustrate the proposed method.
We consider the case for $N=2$, $M=3$ and the number of targets $D=2$.
Suppose the outputs of the first subarray are $\psi_1^{(1)}=-2.6\frac{\pi}{3}$ and $\psi_2^{(1)}=-1.8\frac{\pi}{3}$, and those of the second subarray are $\psi_1^{(2)}=-2.5\frac{\pi}{3}$ and $\psi_2^{(2)}=-1.25\frac{\pi}{3}$. We denote by $\psi_1^{(0)}$ and $\psi_2^{(0)}$ the final results of the estimation.
Using the estimates of the subarrays, we can locate the following four points on the two-dimensional plane:
\textbf{A}$(-2.5\frac{\pi}{3},-2.6\frac{\pi}{3})$, \textbf{B}$(-1.25\frac{\pi}{3},-1.8\frac{\pi}{3})$, \textbf{C}$(-2.5\frac{\pi}{3},-1.8\frac{\pi}{3})$ and \textbf{D}$(-1.25\frac{\pi}{3},-2.6\frac{\pi}{3})$. Also, we can see by simple calculation that
\begin{itemize}
  \item \textbf{A} is closer to L1 with distance $\frac{0.1\pi}{3\sqrt{2}}$;
  \item \textbf{B} is closer to L4 with distance $\frac{0.45\pi}{3\sqrt{2}}$;
  \item \textbf{C} is closer to L3 with distance $\frac{0.3\pi}{3\sqrt{2}}$;
  \item \textbf{D} is closer to L4 with distance $\frac{0.35\pi}{3\sqrt{2}}$.
\end{itemize}
We then project these points back onto the diagonal line segments. For example, \textbf{B} is projected onto L4 and the projected point, say $\textbf{B}_L$, is $(-1.025\frac{\pi}{3},-2.025\frac{\pi}{3})$. Subsequently we construct the modular equations
\begin{align}
\psi^{(0)} \textrm{mod}~ 2 &= -1.025\frac{\pi}{3}\\
\psi^{(0)} \textrm{mod}~ 3 &= -2.025\frac{\pi}{3},
\end{align}
and the solutions will be the value for $\textbf{B}_L$ on the entire angular domain. The equations can be solved using the extended Euclidean algorithm. Alternatively, as we stated before, we can also use a lookup table to solve the equations. After we obtain the four values for the four points, we can then evaluate the likelihood probability at each point and take the largest two values as the final estimates.

\subsection{Discussion}
\subsubsection{Asymptotic performance}
{In this subsection, we briefly discuss the performance of the proposed algorithm. Our method basically consists of two steps. First, we estimate the DOA separately for each subarray, and second, we combine the estimates by projecting the combined point onto a line segment. For the first step, any classical DOA estimation method can be used. In our simulation, we applied the MODE method, which is a type of maximum likelihood estimators \cite{van2002optimum}.}

\begin{figure}[h]
\centering
\includegraphics[scale=0.7, trim=50 380 150 50]{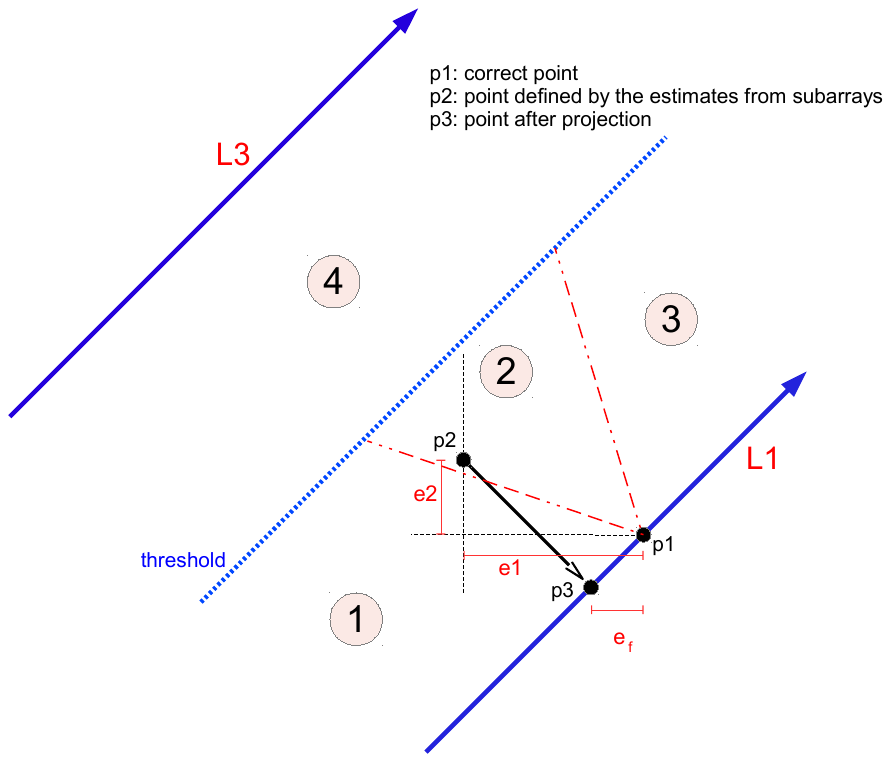}\\
\caption{Illustration of projection error.}\label{figerror}
\end{figure}

{It is well known that maximum likelihood estimators are asymptotically unbiased and normally distributed. The asymptotical property of the proposed algorithm depends on the SNR level. For high SNRs, the point on the plane defined by the two estimates of the subarrays will probably lie very close to the correct point. After projection, it could be even closer. This can be seen on Fig. \ref{figerror}. The estimation errors from the subarrays are indicated by $e_1$ and $e_2$, which moves the combined point $p2$ away from the correct one $p1$. In the second step, we project the point back onto the line segment and obtain $p3$. We can see that the distance between $p1$ and $p3$ is shorter than that of $p1$ and $p2$. It is obvious that as long as $p2$ does not lie outside the blue threshold, which is the line between the two line segments, $p3$ is always closer to $p1$ than $p2$ is to $p1$. Moreover, we notice that the projection is linear. Since the estimates from the subarrays are asymptotically unbiased, we can expect that the final estimate is also unbiased. In other words, the error would not propagate for high SNRs.}

{Under low SNRs, however, $p2$ is more likely to fall on the other side of the blue threshold. As a result, $p2$ will be incorrectly projected to a different line segment and therefore, the final estimate will be poor. This explains why a threshold effect can be seen in the simulations. We point out that this effect is seen in many other unwrapping phase-based methods. After the SNR goes above certain level, the mean square error (MSE) decreases rapidly. We defined four areas in Fig. \ref{figerror}. Given $p1$ as shown in the figure, if the point $p2$ lies in Area 1, we have $e_2<e_f<e_1$; for Area 2, we have $e_f<e_1$ and $e_f<e_2$; for Area 3, we have $e_1<e_f<e_2$;  for Area 4, $p2$ is projected to the other line segment, and therefore the error is magnified. We have  $e_1<e_f$ and $e_2<e_f$.}

\subsubsection{Maximum number of detectable sources}
{For ULAs, it is well known that an array with $N$ elements can identify up to $N-1$ targets. Thus, the two subarrays in the coprime array can identify $N-1$ and $M-1$ targets, respectively. Consequently, there will be $(N-1)(M-1)$ intersections on the two-dimensional plane, which entails that the maximum number of sources that a coprime array can identify is $(N-1)(M-1)$.}

{We point out that our algorithm is particularly suitable for the case of a single target. For multiple sources, it may produce poor results because of the resolution problem in the processing step of each subarray. For ULAs, when two sources are very close to each other, the estimates may be poor. For coprime arrays, it is very likely that the true values of the sources overlap on the wrapped angular domain of one of the subarrays even if on the entire angular domain they do not. In that case, the estimates from the subarray are not accurate at all, and consequently the final estimates are unreliable.}

\section{Simulations}

In this section, we use numerical experiments that demonstrate the performance of the proposed method. We show results of simulations that compare the method with that of the the FD root-MUSIC method as well as with the Cram\'{e}r-Rao bound (CRB). We note that the FD root-MUSIC method uses Fourier series coefficients to approximate the null-spectrum function \cite{rubsamen2009direction}, and we use it because it was shown to have better performance than other search-free methods \cite{gershman2010one}.

In the simulations, we used two sets of arrays, one with $N=5, M=7$, and one with  $N=29, M=31$ elements. For each coprime arrays, we test the cases of one source and three sources.

\subsection{In the case of one source}

First, we simulated one source with $\psi_1=0.1\pi$. In the first two experiments, we studied the performance of the method on the coprime array $N=5, M=7$. In the first of them, we fixed the number of snapshots to $K=100$, and we varied the SNR from $-30$ dB to $10$ dB. For the FD root-MUSIC, we used $S=20, 50, 100,$ and  $200$ Fourier series coefficients, respectively.  We computed the mean square error (MSE) of the methods by using 2000 independent realizations. The results and the CRB are presented in Fig. \ref{fig5}.  We can see that at SNRs below $-8$ dB, the FD root-MUSIC outperforms the proposed method. For SNRs greater than $-8$ dB, the performance of the proposed method approaches the CRB while the FD root-MUSIC suffers from fixed errors due to its approximations. As the number of Fourier series coefficients increases, as expected, the performance of the FD root-MUSIC method improves. However, the method cannot reach the CRB.

In the second experiment, we kept the SNR at $-13$ dB, and varied the number of snapshots from 100 to 1000 in steps of 100. The remaining parameters of the experiment were the same as before. From Fig. \ref{fig6} we see that when the number of snapshots is 100, the MSE of the proposed method is far from the CRB. The results are shown in Fig. \ref{fig6}. We see that the FD root-MUSIC is better than our method when the number of snapshots $K=600$ or smaller. On the other hand, our method approaches the CRB when $K>600$ whereas  the FD root-MUSIC does not improve for $K>400$.

In the remaining two experiments, we repeated the simulations with a coprime array with $N=29, M=31$ elements. Again, first we kept $K=100$ and varied the SNR from $-30$ dB to $10$ dB. The results of the MSE's and the CRB are shown in Fig. \ref{fig7}. We observe similar patterns as in Fig. \ref{fig5}. However, now the proposed method reaches the CRB at $-12$ dB. For SNRs greater than $-12$ dB, the difference in performance between the two methods is by orders of magnitude and in favor of the proposed method.

Finally, in the last experiment, we kept the SNR at $-13$ dB, and varied the number of snapshots from 100 to 1000, in steps of 100. The MSEs of the methods and the CRB are displayed in Fig. \ref{fig8}. We observe, that our method follows the CRB as the number of snapshots increases, whereas the MSE of FD root-MUSIC remains constant.

\ifpfig
\begin{figure}[h]
\centering
\includegraphics[scale=1, trim=250 250 250 260]{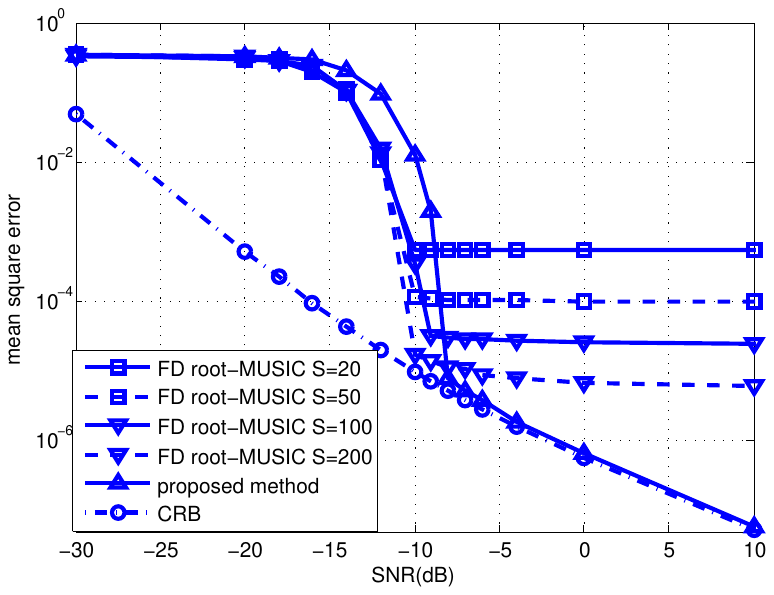}\\ 
\caption{Mean square error performance at different SNRs for one source  ($K$=100). The coprime array is of size (5,7).}\label{fig5}
\end{figure}
\fi

\ifpfig
\begin{figure}[h]
\centering
\includegraphics[scale=1, trim=250 250 250 260]{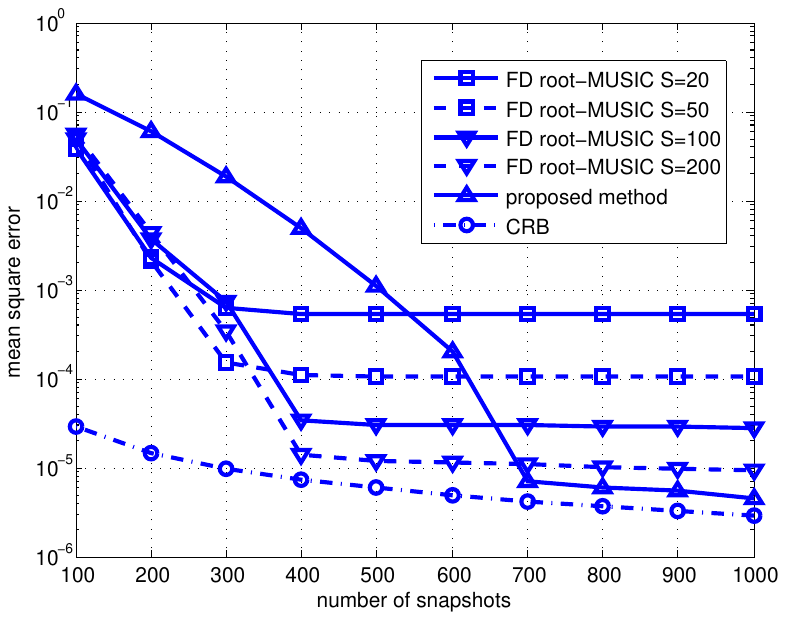}\\ 
\caption{Mean square error performance for different number of snapshots for one source (SNR=$-13$ dB). The coprime array is of size (5,7).}\label{fig6}
\end{figure}
\fi

\ifpfig
\begin{figure}[h]
\centering
\includegraphics[scale=1, trim=250 250 250 260]{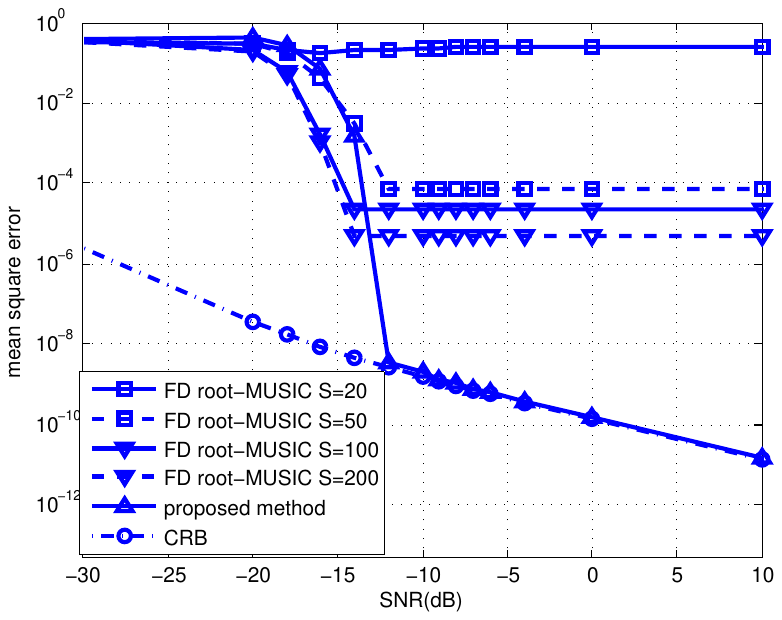}\\ 
\caption{Mean square error performance at different SNRs for one source  ($K$=100). The coprime array is of size (29,31).}\label{fig7}
\end{figure}
\fi

\ifpfig
\begin{figure}[h]
\centering
\includegraphics[scale=1, trim=250 250 250 300]{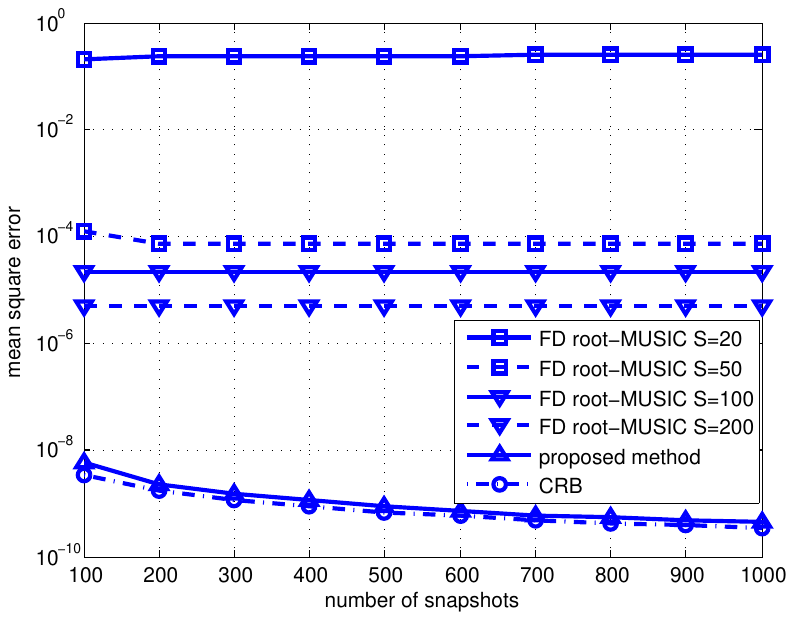}\\
\caption{Mean square error performance for different number of snapshots for one source  (SNR=$-13$dB). The coprime array is of size (29,31).}\label{fig8}
\end{figure}
\fi

\subsection{In the case of three sources}
{In this subsection, we repeated the simulations under the presence of three targets. We set the angles to be $\psi_1=-0.4\pi, \psi_2=0.25\pi, \psi_3=0.33\pi$. Similarly we show four figures here.
The results of the coprime array with $N=5, M=7$ are shown in Fig. \ref{fig9} and Fig. \ref{fig10}.
Fig. \ref{fig9} shows the MSEs for different values of SNR with $K=100$.
We observe similar patterns as in Fig. \ref{fig5} and Fig. \ref{fig7}.
The proposed method reaches the CRB at about $-5$ dB. For SNRs greater than $-7$ dB, the difference in performance between the two methods is by orders of magnitude and in favor of the proposed method. Also we notice that the MSEs of the FD root-MUSIC methods with $S=50,100,200$ are almost the same. That is, with the increase of the approximation order, the performance does not really improve.
Fig. \ref{fig10} shows the MSEs for different values of $K$ with SNR=$-8$ dB.
Our method follows the CRB when the number of snapshots is larger than 400, whereas the MSEs of FD root-MUSIC remains constant.
The results of the coprime array with $N=29, M=30$ are shown in Fig. \ref{fig11} and Fig. \ref{fig12}. Similarly,
Fig. \ref{fig11} shows the MSEs for different values of SNR with $K=100$. The proposed method outperforms the FD root-MUSIC methods under both low and high SNRs. This is because for arrays with large size, 200 Fourier coefficients is far from enough to provide good approximation.
Fig. \ref{fig12} shows the MSEs for different values of $K$ with SNR=$-8$ dB. Our method follows the CRB when the number of snapshots is larger than 900, and its performance is much better than the FD root-MUSIC method with $S=200$.}

{Moreover in Fig. \ref{fig5}, Fig. \ref{fig7}, Fig. \ref{fig9} and Fig. \ref{fig11}, we can observe the threshold effect that we mentioned in Section \ref{section3}. For example, in Fig. \ref{fig7} the MSE of the proposed method rapidly decreases around SNR=$-15$dB; in Fig. \ref{fig9} the threshold effect happens between SNR being $-10$dB and $-5$dB. These justify our discussion in Section \ref{section3}.}

\ifpfig
\begin{figure}[h]
\centering
\includegraphics[scale=1, trim=250 250 250 260]{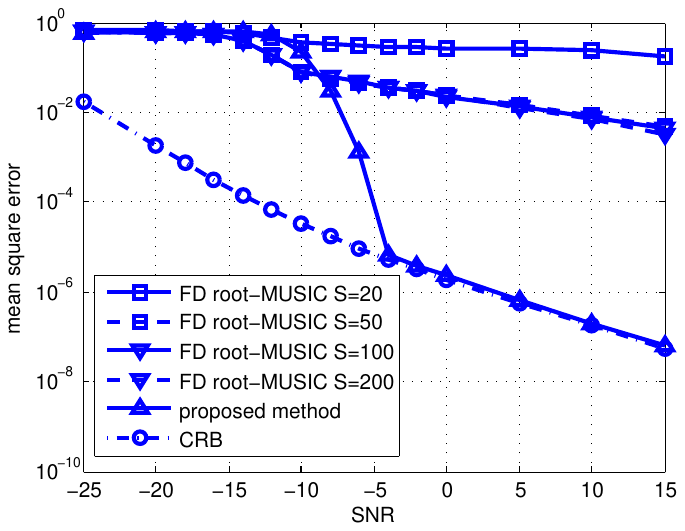}\\ 
\caption{Mean square error performance at different SNRs for three targets ($K$=100). The coprime array is of size (5,7).}\label{fig9}
\end{figure}
\begin{figure}[h]
\centering
\includegraphics[scale=1, trim=250 250 250 300]{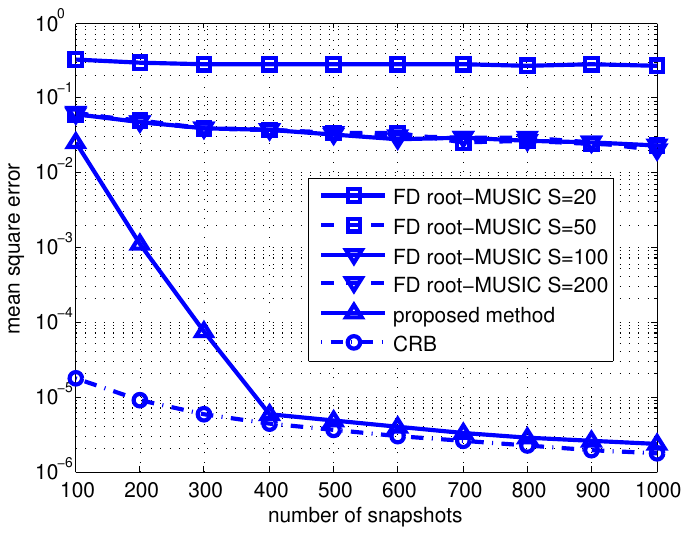}\\
\caption{Mean square error performance for different number of snapshots for three targets (SNR=$-8$dB). The coprime array is of size (5,7).}\label{fig10}
\end{figure}
\fi

\ifpfig
\begin{figure}[h]
\centering
\includegraphics[scale=1, trim=250 250 250 260]{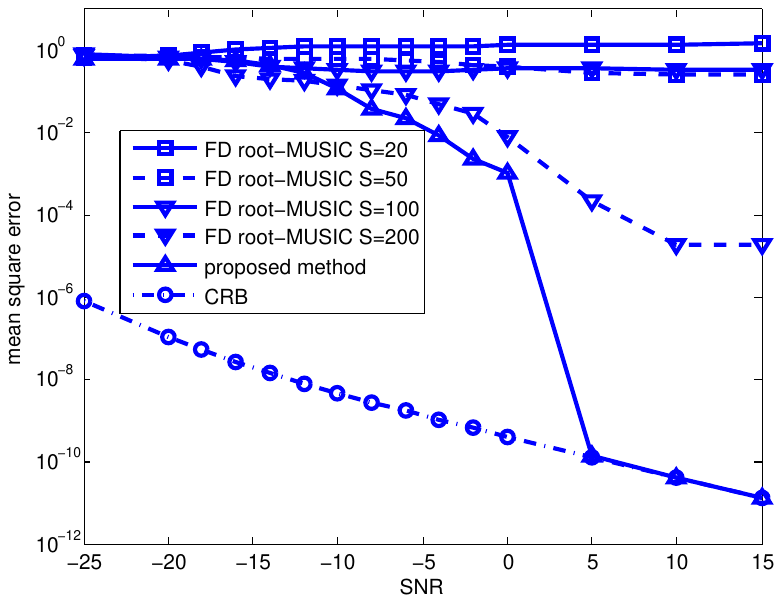}\\ 
\caption{Mean square error performance at different SNRs for three targets ($K$=100). The coprime array is of size (29,31).}\label{fig11}
\end{figure}
\begin{figure}[h]
\centering
\includegraphics[scale=1, trim=250 250 250 300]{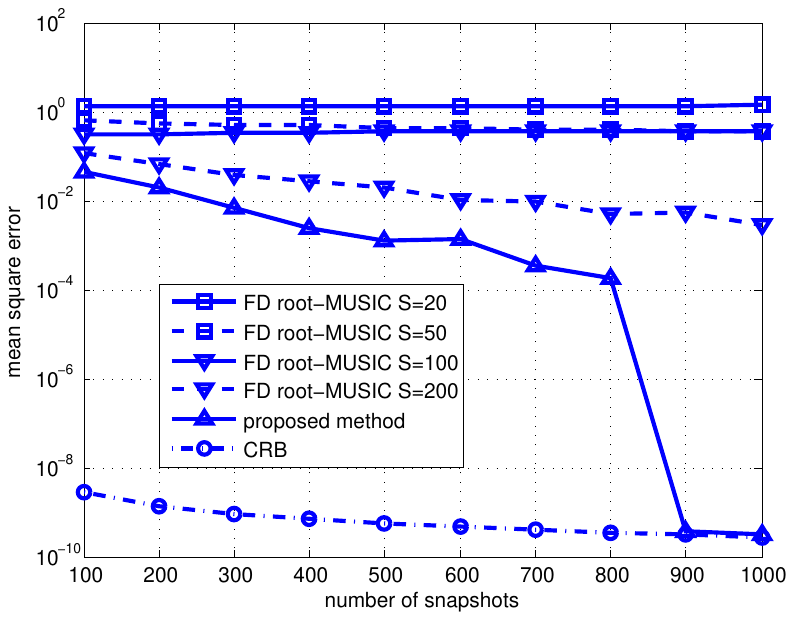}\\
\caption{Mean square error performance for different number of snapshots for three targets (SNR=$-8$dB). The coprime array is of size (29,31).}\label{fig12}
\end{figure}
\fi

In summary, we conclude that the proposed method compares favorably with the FD root-MUSIC method. It is important, however, that we  also make a computational comparison between the methods. A drawback of the FD root-MUSIC method is that it requires the use of a large number of Fourier coefficients to achieve satisfactory performance. In \cite{rubsamen2009direction} it was suggested  that, in general, $S$, the number of Fourier coefficients, should be approximately equal to $4\kappa R$, where $\kappa$ is the signal wavenumber, and $R$ is the largest distance between the array sensors and the origin of the coordinate system. The complexity of FD root-MUSIC is $O((N+M)^2K)$ \cite{rubsamen2009direction}. Thus, to achieve satisfactory performance, the FD root-MUSIC method is computationally expensive even for arrays with small sizes. On the other hand, using the result of MODE in \cite{li1998comparative}, we can obtain that the complexity of the proposed method is $O(M^2K+N^2K)$, which is smaller than that of FD root-MUSIC. 


\section{Concluding remarks}

In this paper, a fast search-free DOA estimation algorithm for coprime arrays is proposed. It exploits the uniform linear structure of the subarrays by first estimating the DOA separately within each subarray and then combining the estimates from different subarrays. Simulation shows that the performance of the proposed algorithm is close to the FD root-MUSIC method at low SNRs and much better at high SNRs. We point out that the proposed method has significantly lower complexity.


Here, we only discussed the case with two subarrays. For a coprime array with more than two subarrays, the state space will be a multi-dimensional hyperrectangle, but the entire angular domain can still be mapped to line segments and the same idea for estimation can be reproduced.  Therefore, the presented results can be extended to coprime arrays with multiple subarrays.


\bibliographystyle{elsarticle-num}
\bibliography{refs}

\section{Vitae}
Zhiyuan Weng was born in Shanghai, China. He is a Ph.D candidate in the Department of Electrical and Computer Engineering at Stony Brook University. His research interests include statistical signal processing and machine learning.

Petar M. Djuri\'{c} received his
B.S. and M.S. degrees in electrical engineering from the University of Belgrade and his Ph.D. degree in electrical engineering
from the University of Rhode Island. Currently, he is a professor
in the Department of Electrical and Computer Engineering at
Stony Brook University. He was an associate editor of several
signal processing journals including IEEE Transactions on Signal Processing. In 2007, he received the IEEE Signal Processing
Magazine Best Paper Award and, during 2008 $-$ 2009, he was Distinguished Lecturer of the IEEE Signal Processing Society. In
2012, he received the European Association for Signal Processing Technical Achievement Award. He is a Fellow of the IEEE.

\end{document}